\documentstyle[prb,preprint,aps]{revtex}
\draft
\input{psfig}
\def\AmS{{\protect\the\textfont2
        A\kern-.1667em\lower.5ex\hbox{M}\kern-.125emS}}

\makeatletter
\def\thepage{1-\@arabic\c@page}
\def\@pnumwidth{2em}
\makeatother
\begin{document}

%\wideabs{
\title{Coupling of carbon nanotubes to metallic contacts}
\author{M. P. Anantram~\cite{byline1}$^1$,  S. Datta$^2$ and 
                                           Yongqiang Xue$^2$}
\address{$^1$ NASA Ames Research Center, Mail Stop T27A-1, Moffett 
              Field, CA, USA 94035-1000  }
\address{$^2$ Department of Electrical and Computer Engineering, 
              Purdue University, West Lafayette, IN 47907-1285}
\maketitle

\begin{abstract}
The modeling of carbon nanotube-metal contacts is important from both
basic and applied view points.
For many applications, it is important to design contacts such that the
transmission is dictated by intrinsic properties of the nanotube rather
than by details of the contact. 
In this paper, we calculate the electron transmission probability from
a nanotube to a free electron metal, which is side-contacted.
If the metal-nanotube interface is sufficiently ordered, we find that
k-vector conservation plays an important role in determining the 
coupling, with the physics depending on the area of contact, tube 
diameter and chirality. The main results of this paper are:
(i) conductance scales with contact length, a phenomena that has been
observed in experiments and
(ii) in the case of uniform coupling between metal and nanotube, the 
threshold value of the metal Fermi wave vector (below which coupling 
is insignificant) depends on chirality.
Disorder and small phase coherence length relax the need for k-vector
conservation, thereby making the coupling stronger.

\vspace{0.3in}

Physical Review B, vol. 61, p. 14219 (2000), APS March 1999 Meeting

\end{abstract}

\section{Introduction} 
\label{section:Introduction}

Carbon nanotubes represent an intriguing new material that has 
attracted much attention both from theorists and experimentalists 
since the early 1990s.~\cite{Dresselhaus_book}
Particularly exciting is the possibility of one dimensional metallic 
conductors at room temperature that can be used as a probe in scanning
probe microscopy or as a low resistance ballistic interconnect for 
electron devices.~\cite{Tans97,White98,Anantram98}
From a more basic point of view, much can be learnt about the physics
of conduction by studying the conductance of such a one dimensional 
conductor at low temperatures.
To exploit these possibilities it is important to understand the
physics of the nanotube-metal contacts and to experimentally
demonstrate low resistance  contacts in a reproducible manner.
The contact between carbon nanotubes and metal can occur at the end of
the tube (end-contact)~\cite{Pablo99,Soh99} and along the circumference
of the tube (side-contact)~\cite{Tans97,Cobden98,Frank98}.
The low contact resistance demonstrated by de Pablo et. 
al~\cite{Pablo99} and Soh et. al~\cite{Soh99} are due to a strong 
interaction between metal and carbon atoms at the end of the nanotube,
or/and due to lack of translational symmetry.~\cite{Tersoff99}
%From a more basic viewpoint, transport through end-contacted nanotubes
%with caps show interesting effects due to localized 
%states.~\cite{Anantram_preprint}
In comparison, the interaction between metal and carbon atoms in 
side-contacted nanotubes is weak.
%The metal can make contact to carbon atoms located either over a 
%sector~\cite{Tans97} or the entire circumference of the 
%nanotube.~\cite{Frank98}

An interesting manifestation of weak distributed coupling is that the
contact resistance is inversely proportional to contact length as 
observed experimentally by references \onlinecite{Tans97} and
\onlinecite{Frank98}. 
Recently, Tersoff in a perceptive paper~\cite{Tersoff99} qualitatively
discussed the importance of k-vector conservation when the coupling
between nanotube and metal is weak. 
The important physical quantities are diameter and chirality of the 
nanotube, Fermi wave vector of the metal, area of contact, and details
of the metal-nanotube contact.
In this paper, we study the physics of side-contacted nanotube-metal 
contacts~\cite{Tans97,Frank98} by addressing how these physical 
quantities affect the transmission of electrons from the nanotube to 
the metal contact.
For small diameter nanotubes, our conclusions do not fully agree with
Ref. \onlinecite{Tersoff99}.
We find that for small diameter armchair tubes, the threshold value of
Fermi wave vector below which the conductance is very small is
$2\pi/3a_0$ and not $4\pi/3a_0$, which is the value for graphene.
$a_0=2.46\AA$ is the lattice vector length of graphene.
In contrast to armchair tubes, the threshold for zigzag tubes is zero.
Our calculations also show that the conductance scales with contact 
length, a phenomena that has been observed experimentally in the work
of Tans et. al~\cite{Tans97} and Frank et. al.~\cite{Frank98}

In the remainder of the introduction, we discuss the salient results 
using simple arguments.
The method is discussed in section \ref{section:method} and the 
numerical results and discussion are presented in 
section \ref{section:results}.
We present our conclusions in section \ref{section:conclusions}.

The first Brillouin zone of graphene touches the Fermi surface at six
points (Fig. 1).
Of these only two points are inequivalent (that is, do not differ by 
a reciprocal lattice vector).
The conduction properties of graphite at low bias are controlled by 
the nature of eigenstates around these points.
Consider a metal making uniform contact to graphene.
The in-plane wave vector should be conserved when an electron tunnels 
from the metal to the nanotube. 
As a result, for good coupling between metal and graphene, the metal
Fermi wave vector should be comparable to $4\pi/3a_0$, which 
corresponds to the Fermi wave vector of graphene.  

To discuss the case of nanotubes making contact to metal, we consider
the scattering rate ($1/\tau_{c-m}$) from the metal to nanotube within
the Born approximation,
\begin{eqnarray}
1/\tau_{c-m} \propto <\Psi_{c}|H_{c-m}|\Psi_m> \mbox{ ,}
                                                  \label{eq:overlap1}
\end{eqnarray}
where, $\Psi_m$ ($\Psi_c$) is the metal (nanotube) wave function and 
$H_{c-m}$ represents the nanotube-metal coupling.
The wave function of an (n,m) nanotube is $\Psi_{c} \sim e^{i k_t p u}
\phi_c$, where $k_t$  is the axial wave vector, $u$ is the 1D unit cell
length, $p$ is an integer representing the various unit cells and 
$\phi_c$ is a vector representing the wave function of all atoms in a
unit cell.
It is assumed that the wave function of the metal is separable in the
axial and radial directions of the nanotube, 
$|\Psi_m> \sim e^{ik_m p u} |\phi_m>$, where $k_m$ is the metal wave 
vector component along the nanotube axis. 
When the coupling between the nanotube and metal is uniform, the 
scattering rate is [Eq. (\ref{eq:overlap1})],
\begin{eqnarray}
1/\tau_{c-m} \propto t_{c-m} <\phi_c|\phi_m> \sum_{p} 
                                e^{i (k_m-k_t) p u}  \mbox{ ,}
\label{eq:goldenrule}
\end{eqnarray}
where, the summation is performed over all unit cells making contact 
to metal and $t_{c-m}$ represents a uniform coupling constant between
the metal and nanotube.
It is clear from Eq. (\ref{eq:goldenrule}) that provided the metal and
nanotube make contact over several unit cells, wave vector conservation
along the axial direction is enforced as
$\sum_{p} e^{i (k_m-k_t) p u} \sim \frac{1}{u} \delta(k_m-k_t)$.
The axial wave vector corresponding to $E=0$  are $2\pi/3a_0$ and $0$
for armchair and zigzag tubes respectively, and the wave vector for 
other chiralities varies between these two limits.
As a result, the threshold value of Fermi wave vector below which 
coupling between an armchair (zigzag) nanotube and metal is poor is 
$2\pi/3a_0$ ($0$).
The threshold value of the metal Fermi wave vector for chiral tubes is
in between that of zigzag and armchair tubes.
As the diameter of the nanotube increases, wave vector conservation 
along the circumference becomes increasingly important, as the strip 
approaches a graphene sheet.

\section{Method}
\label{section:method}

The method used to calculate transmission probability is essentially 
the same as that in reference \onlinecite{Anantram98}, with the only 
addition being the connection of a metal contact.~\cite{Caroli71}
So in this section, we mainly focus on connection to the metal contact.
The metal contact has a rectangular cross section in the (x,z) plane 
and is infinitely long along the y-axis as shown in Fig. 2.
The nanotube lies on the metal contact akin to the experiment of Tans
et.  al~\cite{Tans97}.
In reference \onlinecite{Tans97}, the nanotube bends over the edge of
the metal and the influence of this on transport has recently been 
modeled by Rochefort et. al.~\cite{Rochefort99}.
In this work, the main focus is to model the coupling between the metal
and nanotube. So we assume the nanotube to lie rigidly on the metal and
neglect the effect of bending (Fig. 2).
A perfectly cylindrical nanotube would touch the metal surface only 
along a line.  To simplify modeling this interface, we stretch the 
entire circumference of the nanotube over the metal surface and assume
coupling between carbon atoms in a sector of the circumference and the
metal.  Finally, charge self consistency~\cite{Odintsov99} has been 
neglected.

The transmission and local density of states are calculated in a 
structure that can be conceptually divided into four parts: section of
the nanotube (D), which lies on the metal electrode (M), semi-infinite
regions of the nanotube L and R [Fig. 2]. The Hamiltonian of the 
system can be written as,
\begin{eqnarray}
H = H_c + H_m + H_{c-m} \mbox{ and } \\
H_c = H_D + H_L + H_R + H_{LD} + H_{RD}
\end{eqnarray}
where, $H_c$ is the pi-electron tight binding Hamiltonian of the 
nanotube with the on-site potential and hopping parameter between 
nearest neighbor carbon atoms equal to 0 and 3.1 eV 
respectively.~\cite{Dresselhaus_book}
$H_{LD}$ and $H_{RD}$ are terms in the Hamiltonian coupling D to L and
R respectively.  $H_m$ and $H_{c-m}$ are the free particle and 
nanotube-metal coupling terms of the Hamiltonian. The Green's function
$G^r$ is obtained by solving:
$\left[ E - H_D - \Sigma^r_L - \Sigma^r_R - \Sigma^r_m \right] G^r(E)
 = I$, where the self energy 
$\Sigma_\alpha = V_{D \alpha} \; g_\alpha^r \; V_{\alpha D}$
($\alpha \; \in$ {L}, {R} and {M}).
$g_\alpha^r$ is the surface Green's function of lead $\alpha$ and
$V_{D \alpha}$ ($V_{\alpha D}$) is the coupling between $D$ ($\alpha$)
and $\alpha$ ($D$). The transmission probability between leads $\alpha$
and $\beta$ [$T_{\alpha\beta}$] is given by,
\begin{eqnarray}
T_{\alpha \beta}(E) = Trace[\Gamma_{\alpha}(E)  G^r(E) 
                                   \Gamma_{\beta}(E) G^a(E)] \mbox{ ,}      
						\label{eq:Transmission}
\end{eqnarray}
where $\Gamma_\alpha(E) = 2 \pi V_{D \alpha} \; \rho_\alpha(E) \; 
V_{\alpha D}$ and $\rho_\alpha(E) = -\frac{1}{\pi} Im[g_\alpha^r(E)]$
is the surface density of states of lead $\alpha$.

The Green's function of the metal contact is calculated within the free
electron approximation using the procedure outlined below. The metal 
contact has a rectangular cross section of dimensions $L_x$ and $L_z$
in the x and z directions respectively, and is infinitely long in the
y direction.
While the (y,z)-coordinates are assumed to be continuous, the 
x-coordinate is assumed to be discrete with lattice spacing
$a=L_x/(N_x+1)$, where $N_x$ is the number of lattice points. The wave
functions ($\Psi_{mkn}$) and eigen values ($E_{mkn}$) are given by,
\begin{eqnarray}
\Psi_{mkn} (r) &=& X_m(x) Y_k(y) Z_n(z) \mbox{ , where,}
					  \label{eq:wavefun} \\
X_m(x) &=& \frac{1}{\sqrt{L_x}} sin (\frac{m \pi x}{L_x}) \mbox{ , }
%\nonumber\\
Y_k(y) = \frac{1}{\sqrt{L_y}} exp(iky)  \mbox{ , }
%              \nonumber\\
Z_n(z) = \frac{1}{\sqrt{L_z}} sin (\frac{m \pi z}{L_z}) \mbox{ and }
\nonumber \\
E_{mkn} &=& \frac{\hbar^2}{2m_oa^2}
[1-cos(\frac{m\pi}{N_x+1})]+\frac{\hbar^2k^2}{2m_o} +
\frac{\hbar^2}{2m_o} (\frac{n\pi}{L_z})^2 \mbox{ ,} \label{eq:energy}
\end{eqnarray}
where, $m$ and $n$ are positive integers, and $m_0$ is the free
electron mass. Using Eqns. (\ref{eq:wavefun}) and (\ref{eq:energy}) in
the following equation for the Green's function,
\begin{eqnarray}
g(r,r^\prime,E) = \sum_{m,k,n} \frac{\Psi_{mkn}(r)^\ast
\Psi_{mkn}(r^\prime)}{E-E_{mkn}+i\eta} \mbox{ ,} \nonumber
\end{eqnarray}
we obtain,
\begin{eqnarray}
g(r,r^\prime,E) = -\frac{im_o}{\hbar^2} \frac{1}{L_xL_z} \sum_{m,n}
\frac{exp[ik_I|y-y^\prime|]}{k_I} sin(\frac{m\pi x}{L_x})
sin(\frac{m\pi x^\prime}{L_x}) sin(\frac{n\pi z}{L_z}) 
sin(\frac{n\pi z^\prime}{L_z})
\mbox{ ,} \label{eq:g_M}
\end{eqnarray}
where,
\begin{eqnarray}
k_I = \{k^2 - (\frac{n\pi}{L_z})^2 - \frac{1}{a^2}
[1-cos(\frac{m\pi}{N_x+1})] + i\eta\}^{\frac{1}{2}} \mbox{ and }
k=\sqrt{\frac{2m_oE}{\hbar^2}} \mbox{.} \label{eq:k}
\end{eqnarray}
For carbon nanotubes, the zero of energy ($E=0$) is taken to lie at
the band center. 
On the other hand, in deriving Eq. (\ref{eq:g_M}) the zero of energy
corresponded to the band bottom of the free electron metal.
In the calculations, there should be only one zero of energy, which we
take to lie at the band center of the nanotube.
We also neglect charging effects and assume the Fermi energy of the
metal to lie at the band center of the nanotube.~\cite{footnote2}
Then, in the coordinate system where $E=0$ corresponds to the band
center of the nanotube, Eq. (\ref{eq:g_M}) can be used by transforming,
\begin{eqnarray}
k=\sqrt{\frac{2m_oE}{\hbar^2}} \;\;\;\;\; \mbox{ to } \;\;\;\;\;
k=\sqrt{\frac{2m_oE}{\hbar^2} + k_f^2} \;\;\;\;\;
                                    \mbox{ in Eq. (\ref{eq:k}),} 
\end{eqnarray}
where, $k_f$ is the Fermi wave vector of the metal.

The component of the Green's function that enters the calculation of 
the density of states and transmission probability corresponds to 
$x=x^\prime=a$, the surface of the metal contact on which the nanotube
lies. The $(y,z)$ coordinates correspond to the atomic location of the
stretched out nanotube lying on the metal. For uniform coupling between
the metal and nanotube, we take $V_{DM}=tD_0$, where $t$ is the 
strength of coupling between the free electron metal and a nanotube 
atom and $D_0$ is a diagonal matrix whose dimension is equal to the
number of carbon atoms in $D$. The diagonal entry $D_0(i,i) = 1 (0)$
if the carbon atoms make (do not make) contact to the metal.

\section{Results and Discussion}
\label{section:results}

We first present results for dependence of the threshold value of the
metal Fermi wave vector on chirality, using armchair and zigzag tubes
connected to the metal contact. We then discuss the diameter dependence
of conductance using the case of a zigzag tube as an example. Finally,
the case of disorder in coupling between the nanotube and metal is 
considered. We consider only weak coupling between the nanotube and 
metal. The average value of the non zero diagonal elements of the 
coupling strength $\Gamma_M$ are tabulated in Table 1 for the various
values of the metal Fermi wave vector considered.
The main guide for the choice of $\Gamma_M$ is that it be much smaller
than the corresponding coupling strength between two carbon atoms of 
the nanotube (diagonal component of $\Gamma_L$ is approximately equal
to $0.3eV$ for a (2,2) nanotube).
A larger/smaller value of $\Gamma_M$ results in a larger/smaller
value of transmission in Figures 2 to 5. We calculate the transmission
versus contact length between nanotube and metal for various Fermi wave
vectors in the metal and all atoms around the circumference of the tube
are assumed to make uniform contact with the metal. 
We emphasize that when the metal makes contact to only a sector of the
nanotube such as in Ref. \onlinecite{Tans97}, the results of Fermi wave
vector dependence on chirality and the conductance dependence on
contact length are still valid. These features depend on the 
nanotube-metal coupling along the axial direction. So any change due to
the finite sector will not qualitatively change the results.

Experiments typically involve transmission of electrons between two
metal contacts. The quantity $T_{ML}$ discussed in this section is
however the transmission probability between a metal contact and a
semi-infinte nanotube (Fig. 2). We consider this quantity because a 
long nanotube section between two metal contacts requires much more
numerically intensive calculations. The physics discussed with regards
to $T_{ML}$ in Figs. 3-5 hold in the case of two metallic contacts
also, though a direct numerical comparison is not appropriate.

In the case of armchair tubes, when the metal Fermi wave vector $k_f$
is smaller than $2\pi/3a_0$ ($0.85 \AA^{-1}$), $T_{ML}$ does not change
significantly with contact length as shown for $k_f =0.75 \AA^{-1}$ in
Fig. 3(a). For values of $k_f$ above the threshold, the transmission 
monotonously increases with an increase in contact length. The 
monotonic increase is due to weak metal-nanotube coupling, in which 
case an increase in contact length simply results in an increase in 
the transition probability to scatter from metal to 
nanotube.~\cite{footnote1} The transmission will eventually saturate 
with increase in contact length as there are only two conducting modes
at the band center. For the configuration considered, $T_{ML}$ can have
a maximum value of unity.  The second feature of Fig. 3(a) is the 
increase in transmission with increase in $k_f$. This can be understood
by noting that electrons with a wave vector component along the 
nanotube axis that is larger than $2\pi/3a_0$ scatter from the metal 
to nanotube, and a larger $k_f$ implies a large number of available 
metal electron states. For the purpose of these calculations, we 
considered a (2,2) armchair tube; The essential physics would in 
principle be true for the  more realistic (10,10) nanotube also.

The case of zigzag tubes is different because bands at $E=0$ cross at
$k=0$. Then, electrons in the metal electrode with any $k_f$ (no 
threshold) can scatter into a metallic zigzag tube. The results for a
(3,0) tube are shown in Fig. 3(b). Here, there are two important 
points. The first point is that as there is no threshold metal Fermi 
wave vector, the transmission increases monotonically with contact 
length even for $k_f=0.4 \AA^{-1}$, which is smaller than the threshold
for armchair tubes. The second point is that the transmission for $k_f$
equal to $1.2\AA^{-1}$ is much smaller than that for armchair tubes 
[Fig. 3(a); the transmission of the three smaller values of $k_f$ have
been multiplied by a factor of ten.]. This is because the nanotube wave
vector around the circumference ($k_c$) of a zigzag tube is large;
$k_c=4\pi/3a_0$ for the crossing bands and as a result, the overlap 
integral [Eq. (\ref{eq:overlap1})] is small. As $k_f=1.75\AA^{-1}$ is
larger than the threshold for graphite, the transmission probability is
larger, and comparable to that for armchair tubes [Fig. 3(b)].

What happens when the diameter increases?
In the limit of large diameter, a nanotube is akin to graphene and the
threshold $k_f$ to couple well with metal should approach
$4\pi/3a_0$.~\cite{Tersoff99} Numerically, it is difficult to simulate
large diameter tubes along with large contact lengths because of time
and memory requirements associated with the calculation of $g_M^r$. To
convey the main point we consider two simpler cases, the first case 
compares the transmission probability of the two smallest semi-metallic
zigzag tubes with varying contact lengths and the second case considers
zigzag tubes of varying diameters with a rather small contact length.
Fig. 4 compares the transmission probability versus contact length of
the (3,0) and (6,0) nanotubes; The (6,0) nanotube has double the 
diameter of the (3,0) nanotube. The (6,0) nanotube correspondingly has
a smaller transmission and the trend of decrease in transmission will
continue with further increase in diameter. 
The inset is a calculation of transmission probability versus diameter
of semi-metallic zigzag tubes for a contact length of 42.6 $\AA$ (ten
unit cells). 
$T_{ML}$ decreases with increase in diameter because wave vector 
conservation becomes increasingly important with increase in diameter.
Shown also in this figure for comparison are $1/\mbox{diameter}$ and 
$1/\sqrt{\mbox{diameter}}$.

We now address the role of disorder.
Disorder in either the nanotube, metal or nanotube-metal coupling will
in general result in larger transmission when compared to the 
disorder-free case. Wave vector conservation is relaxed due to 
scattering from defects and transmission will increase with increase 
in contact length {\it even when the metal $k_f$ is below the threshold
value}. We consider the case of disorder in nanotube-metal coupling 
($H_{c-m}$). Disorder in all elements of the coupling between the 
nanotube and metal was introduced randomly. The disorder in coupling 
of atom $i$ to the metal contact can be written as, 
$t_i = \alpha t^{av} + (1-\alpha) t_i^{rand}$, where $t^{av}$ is the 
average value of $t_i$ over all sites connected to the metal and 
$\alpha$ is a fraction between zero and unity. $t_i^{rand}$ is the 
random component whose average is equal to $t^{av}$. In Fig. 5, the two
strengths of disorder correspond to $\alpha=0$ and $\alpha=0.5$ 
(smaller $\alpha$ corresponds to larger disorder), such that $t^{av}$
has the same value as in Fig. 3(a). For an armchair tube in contact 
with a metal with $k_f =0.75 \AA^{-1}$, the transmission was very small
and more importantly did not vary with contact length [Fig. 3(a)].
Introducing disorder changes this trend and causes a monotonic increase
in transmission with length of contact [Fig. 5]. Similarly, for large
diameter tubes, in the presence of disorder there should be significant
transmission when $k_f$  is smaller than the threshold $4\pi/3a_0$. The
requirement of wave vector conservation is also relaxed when the phase
coherence length is small. So we expect the coupling to improve with 
decrease in phase coherence length.

\section{Conclusions}
\label{section:conclusions}

In this paper, we addressed some aspects of the physics of a nanotube
side-contacted to metal, a problem of current importance. Coupling of
carbon nanotubes to metal depends on both chirality and diameter. Wave
vector conservation of an electron scattered from the nanotube to metal
plays a central role in determining the transport properties. The 
difference between small and large diameter nanotubes is that while in
the former wave vector conservation is important only in the axial 
direction, in the latter it is important in both the axial and
circumferential directions. As a result, small diameter armchair and 
zigzag tubes have a cut-off value of the metal Fermi wave vector equal
to $2\pi/3a_0$  and zero, respectively. For chiral tubes, the cut-off
value of the metal Fermi wave vector lies in between these two limits,
with the value decreasing with increase in chiral angle. A large 
diameter nanotube is akin to a graphene sheet and the cut-off value of
the metal Fermi wave vector in this case approaches $4\pi/3a_0$ with 
increase in diameter. Disorder in the metal, nanotube or their coupling
relaxes the requirement of k-vector conservation and in general 
improves coupling. The groups of references \onlinecite{Tans97} and 
\onlinecite{Frank98} have shown increase in conductance with contact 
length.  In this paper, we discussed two situations that could lead to
this. The first situation requires the metal Fermi wave vector to be 
larger than the threshold discussed in the text and holds even when 
there is no disorder. The second situation requires disorder in 
coupling to the metal but there is no restriction on the value of the
Fermi wave vector.

\section{Acknowledgements}
\label{acknowledgements}

We acknowledge useful discussion with W. A. de Heer (Georgia Tech), 
Cees Dekker and Zhen Yao (both of Delft University) and thank 
J. Tersoff (IBM) for providing us with a preprint of reference 
\onlinecite{Tersoff99}. We thank Mario Encinosa (FAMU) for many useful
comments on the manuscript, and Alexei Svizhenko (NASA Ames) for help
with commands to parallelize code.

\pagebreak

\noindent
{\bf Figure Captions:}

\noindent
Fig. 1: First Brillouin zone of graphene. Points $P$, $P^\prime$,
$P^{\prime\prime}$, $Q$, $Q^\prime$, $Q^{\prime\prime}$  touch the
Fermi surface.  $a_0$ is the lattice vector length of graphene.
A metal with Fermi wave vector smaller (inner circle) and larger (outer
circle) than $4\pi/3a_0$ couples poorly and well to graphene
respectively. $\Delta E_{NC}$ is the energy difference between the
first non crossing subband below and above E=0.

\vspace{0.1in}

\noindent
Fig. 2: A metal making contact to a nanotube.
The $(x,z)$ dimensions of the metal form a rectangular cross section 
with lengths $(L_x,L_z)$. The $y$ direction is infinitely long.

\vspace{0.1in}

\noindent
Fig. 3: Transmission probability for (a) armchair and (b) zigzag tubes
versus contact length. In both cases the largest contact length 
corresponds to sixty unit cells.
The main point of (a) is that for the metal Fermi wave vector smaller
than the threshold $2\pi/3a_0$, coupling between the nanotube and metal
is small and increasing the contact length does not change the 
transmission probability. For metal Fermi wave vector larger than 
$2\pi/3a_0$, the transmission probability increases with increase in 
contact length and also with increase in $k_f$ for a given contact 
length.  The main point of (b) is that there is no threshold in the 
metal Fermi wave vector.
Even in the case of a small value of the metal Fermi wave vector
($0.4\AA^{-1}$), the transmission increases with increase in the 
contact length, albeit the magnitude of transmission is small.
As in the armchair case, the transmission probability increases with
increase in $k_f$ for a given contact length.
The values of $T_{ML}$ in (b) corresponding to $k_f$ equal to 0.4,
0.75 and 1.2$\AA$ are multiplied by a factor of ten.

\vspace{0.1in}

\noindent
Fig. 4:  Comparison of transmission probability of (3,0) and (6,0)
nanotubes versus contact length.
The transmission probability decreases with increase in diameter.
Inset: The y-axis is $T_{ML}$ for metallic zigzag tubes scaled by 1.0
e+4. The solid line is the diameter dependence of $T_{ML}$ for a
contact length of 42.6 $\AA$. The upper and lower dashed lines are
$1/\sqrt{\mbox{diameter}}$ and $1/\mbox{diameter}$ dependences,
shown for comparison.

\vspace{0.1in}

\noindent
Fig. 5: Comparison of transmission probability versus contact length 
for a (2,2) armchair tube, with and without disorder in nanotube-metal
coupling.  The metal Fermi wave vector is $0.75 \AA^{-1}$. Note that 
for the case without disorder, the transmission is poor and increasing
the contact length does not help. Introducing disorder changes this 
picture and the transmission begins to increase with increase in 
contact length because k-vector conservation is relaxed.

\pagebreak

\begin{table}[h]
\centerline{\psfig{file=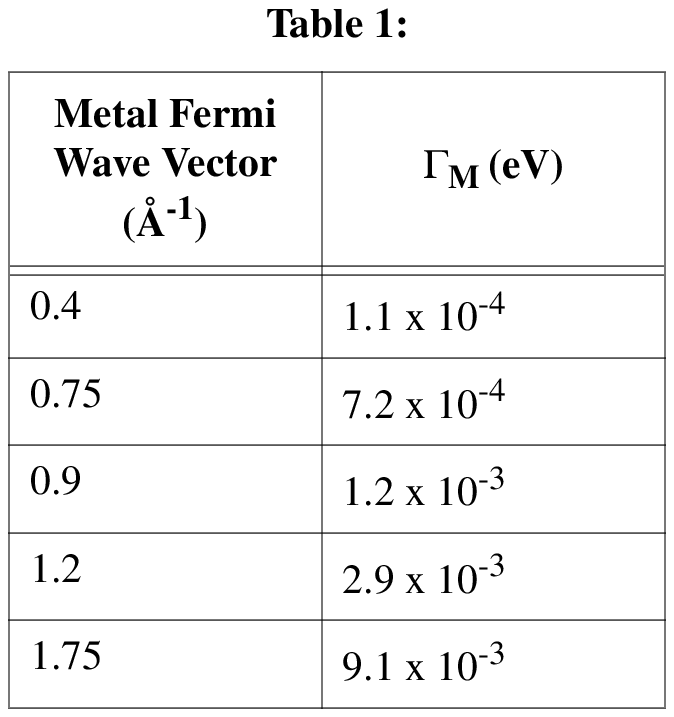}}
\small $\Gamma_M$ for the different values of the metal Fermi wave 
vectors used.
\end{table}

\pagebreak

\begin{figure}[h]
\centerline{\psfig{file=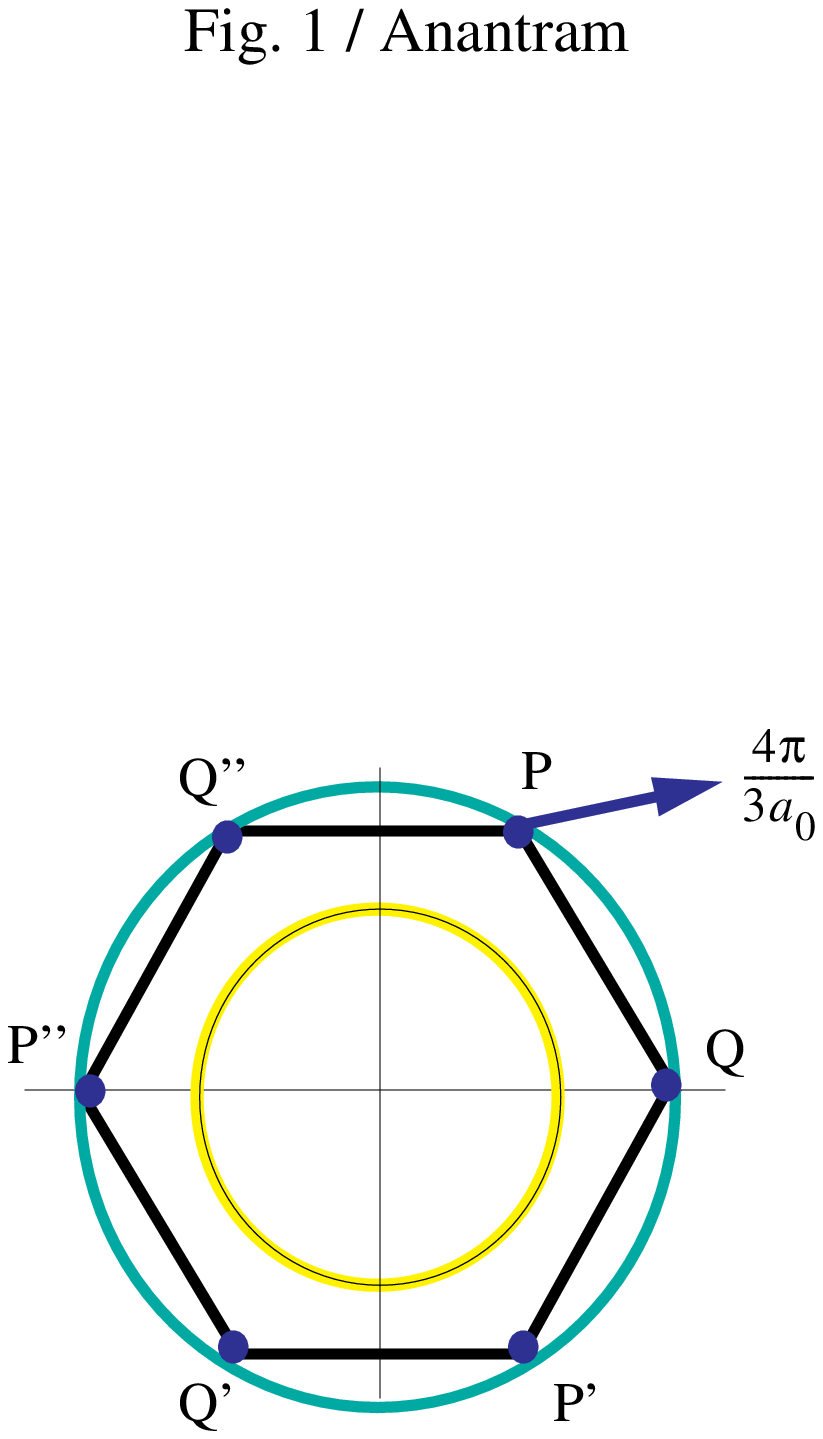}}
\small
\end{figure}

\pagebreak

\begin{figure}[h]
\centerline{\psfig{file=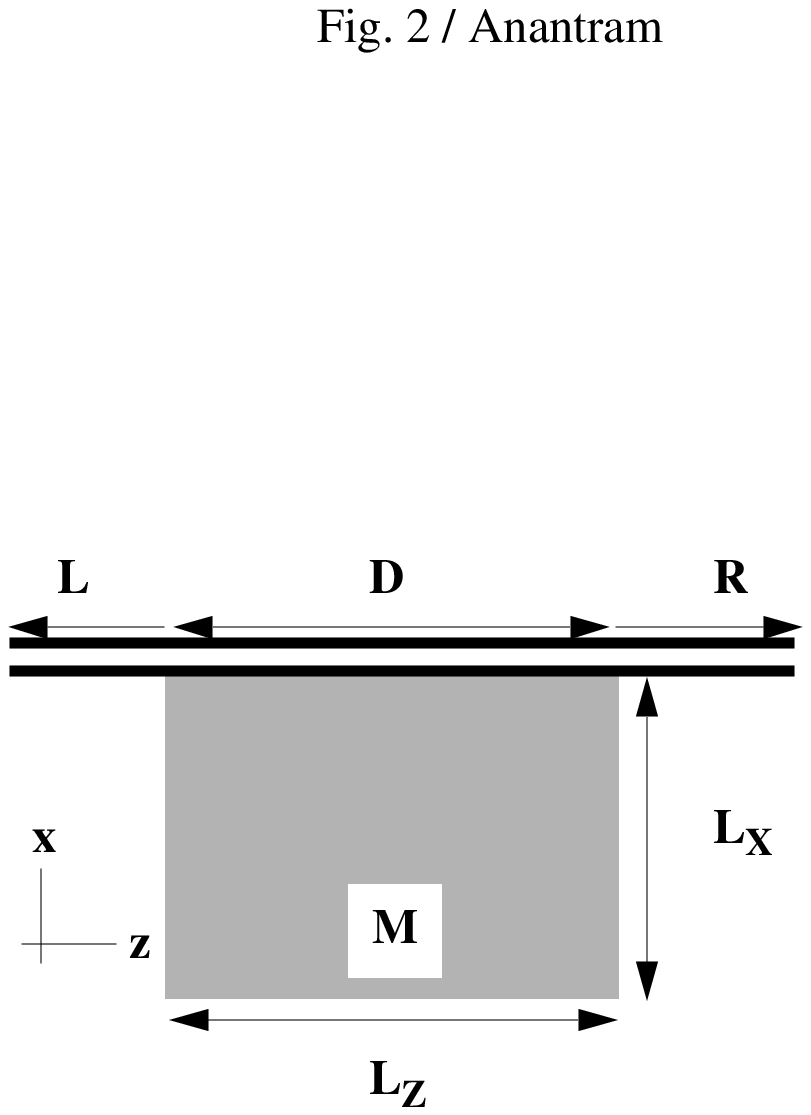}}
\small
\end{figure}

\pagebreak

\begin{figure}[h]
\centerline{\psfig{file=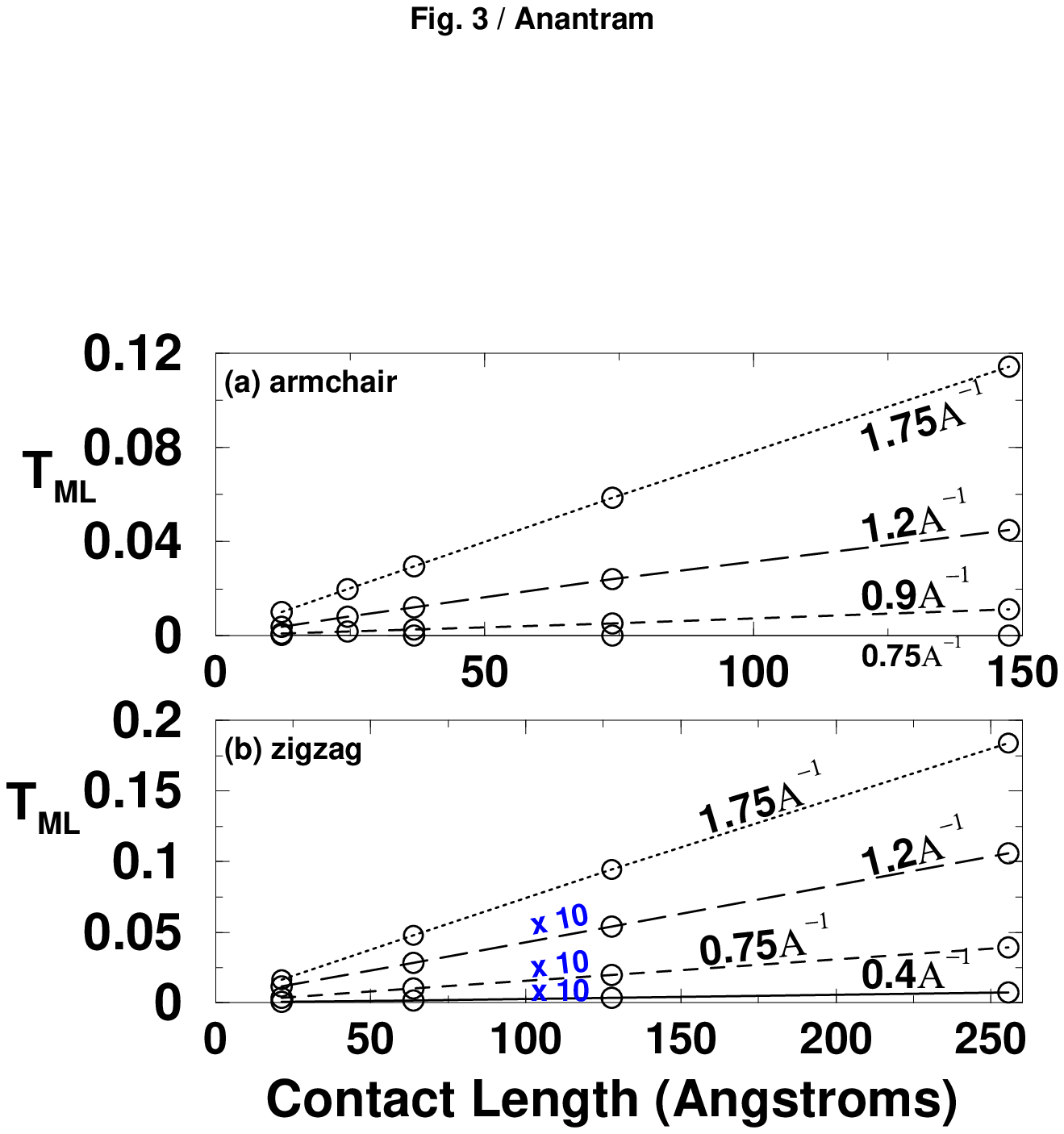}}
\small
\end{figure}

\begin{figure}[h]
\centerline{\psfig{file=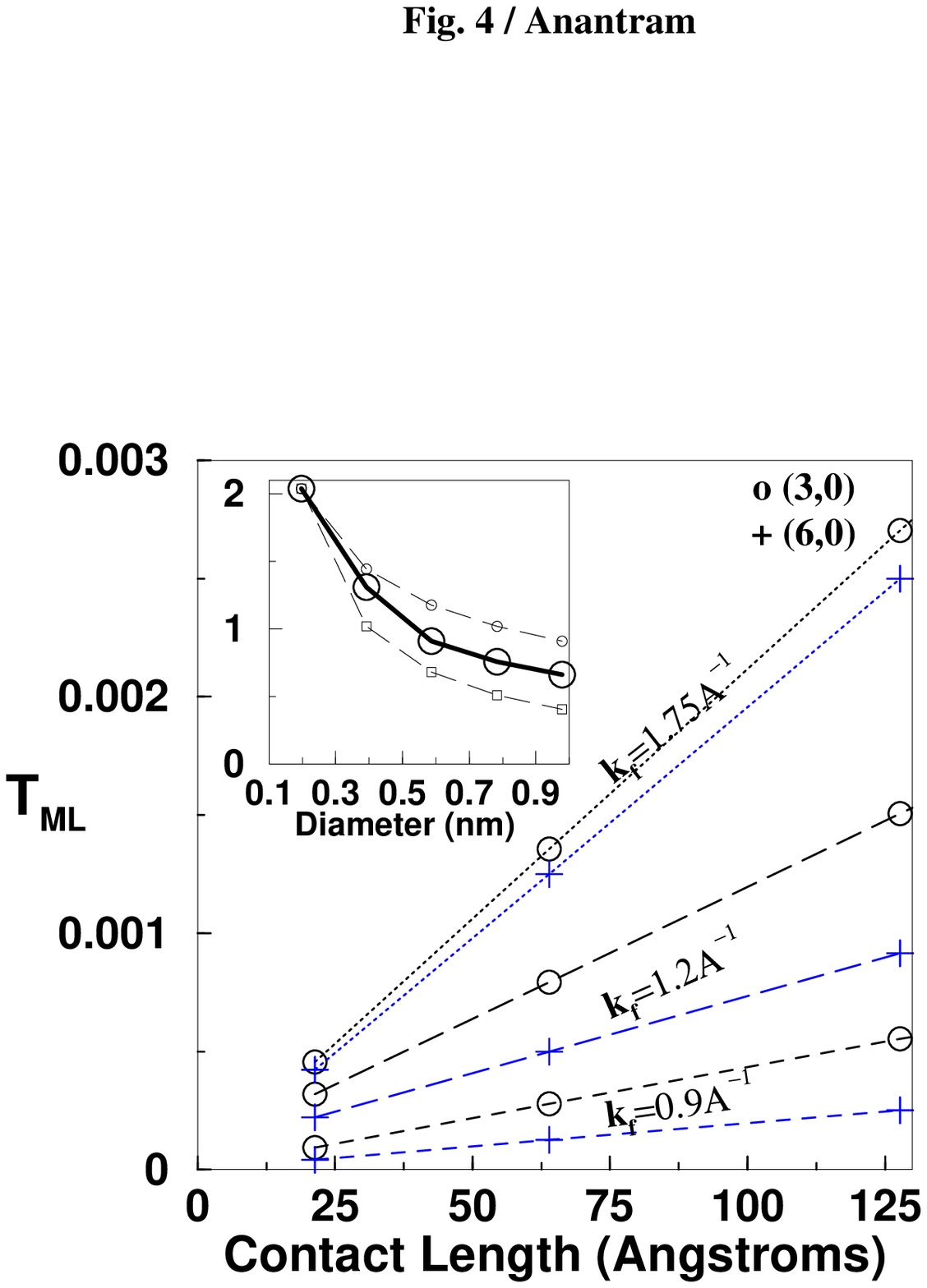}}
\small
\end{figure}

\pagebreak

\begin{figure}[h]
\centerline{\psfig{file=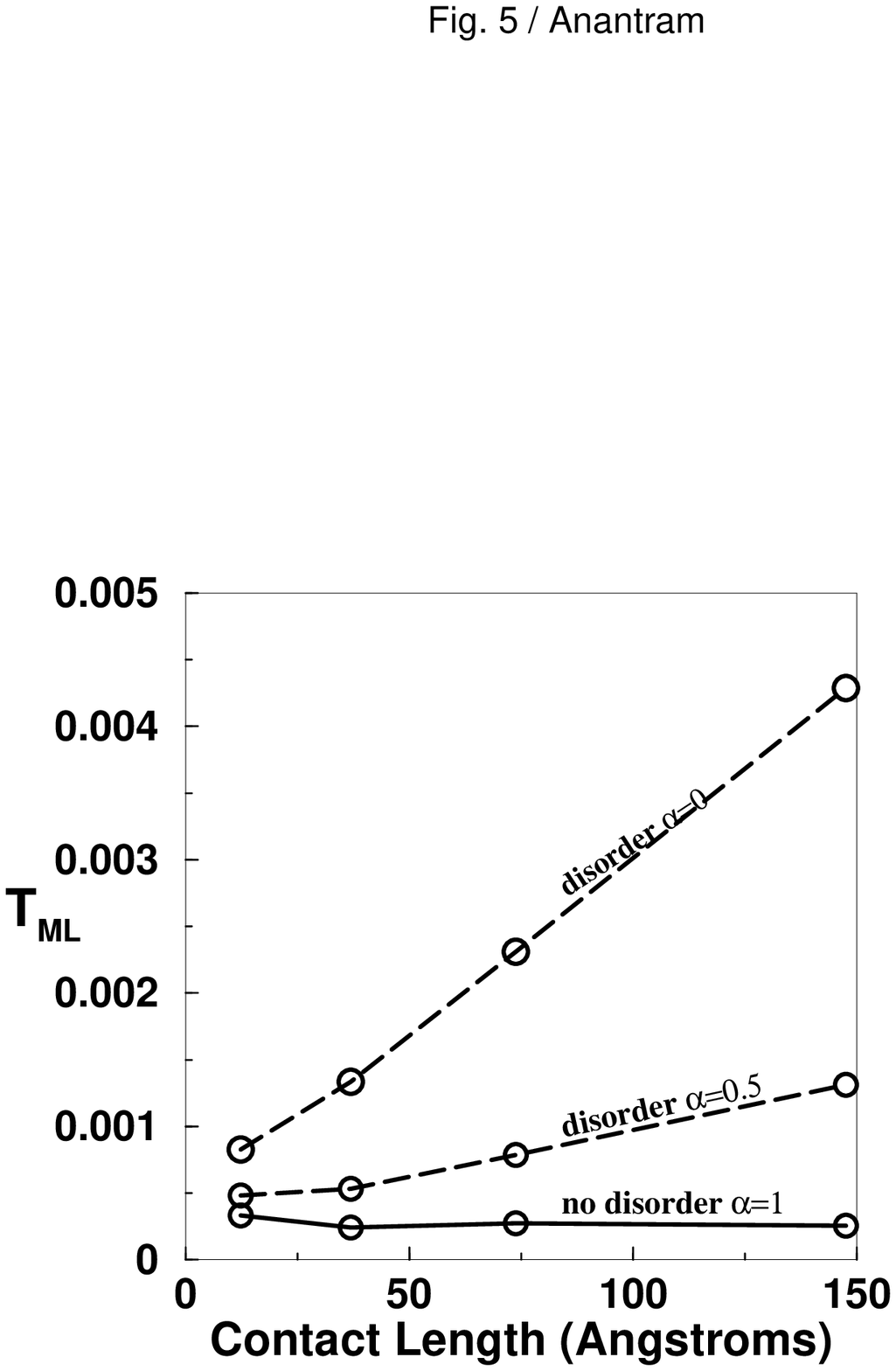}}
\small
\end{figure}


\begin{thebibliography}{100}

\bibitem[*]{byline1} Corresponding author: anant@nas.nasa.gov.

\bibitem{Dresselhaus_book}
M. S. Dresselhaus, G. Dresselhaus and P. C. Eklund,
Chap. 19 of Science of Fullerenes and Carbon Nanotubes, Academic Press,
(1996)

\bibitem{Tans97}
S. J. Tans, M. Devoret, H. Dai, A. Thess, R.E. Smalley, L.J.
Geerligs and C. Dekker,
%Individual single-wall carbon nanotubes as quantum wires.
Nature {\bf 386}, 474 (1997).


\bibitem{White98}
C. T. White and T. N. Todorov,
Nature {\bf 393}, 240 (1998)

\bibitem{Anantram98}
M. P. Anantram and T. R. Govindan,
%Conductance of carbon nanotubes with disorder: A numerical study
Phys. Rev. B {\bf 58}, 4882 (1998)

%\bibitem{Dai96}
%H. Dai et. al,,
%J. H. Hafner, A. G. Rinzler, D. T. Colbert and R. E. Smalley,
%Nature {\bf 384}, 147 (1996)

%\bibitem{Ebbesen96}
%T. W. Ebbessen et. al,
%H. J. Lezec, H. Hiura, J. W. Bennett, H. F. Ghaemi and T. Thio
%Nature {\bf 382}, 54 (1996)

%\bibitem{Bezryadin98}
%A. Bezryadin et. al,
% A. R. M. Verschueren, S. J. Tans and C. Dekker,
%Multiprobe Transport Experiments on Individual Single-Wall 
%Carbon Nanotubes
%Phys. Rev. Lett {\bf 80}, 4036 (1998)

\bibitem{Pablo99}
P. J. de Pablo,
E. Graugnard, B. Walsh, R. P. Andres, S. Datta and R. Reifenberger,
Appl. Phys. Lett. {\bf 74}, 323 (1999)

\bibitem{Soh99}
H. T. Soh, A. F. Morpurgo, J. Kong, C. M. Marcus, C. F. Quate and 
H. Dai, Preprint (1999).

\bibitem{Cobden98}
D. H. Cobden,
M. Bockrath, P. L. McEuen, A. G. Rinzler and R. E. Smalley,
Spin Splitting and Even-Odd Effects in Carbon Nanotubes
Phys. Rev. Lett {\bf 81}, 681 (1998)


\bibitem{Frank98}
S. Frank,
P. Poncharal, Z. L. Wang and W. A. de Heer,
Science {\bf 280}, 1744 (1998); P. Poncharal, S. Frank, Z. L. Wang and
W. A. de Heer, Conductance quantization in multiwalled carbon nanotubes
(Preprint)

\bibitem{Tersoff99}
J. Tersoff,
%Contact resistance of carbon nanotubes
Appl. Phys. Lett. {\bf 74}, 2122 (1999)

%\bibitem{Anantram_preprint}
%M. P. Anantram and T. R. Govindan,
%{\it Transmission through capped carbon nanotubes with polyhedral
%caps},
%cond-mat/9907020

\bibitem{Caroli71}
The following references discuss the Green's function formalism used:
S. Datta, Electronic Transport in Mesoscopic Systems,
Cambridge University Press, Cambridge, U.K (1995);
C. Caroli, R. Combescot, P. Nozieres and D. Saint-James,
%Direct calculation of the tunneling current.
J. Phys. C: Solid St. Phys. {\bf 4}, 916 (1971);
Y. Meir and N.S. Wingreen,
%Landauer Formula for the Current through an Interacting
%Electron Region.
Phys. Rev. Lett. {\bf 68}, 2512 (1992)

\bibitem{Rochefort99}
A. Rochefort, F. Lesage, D.R. Salahub, and Ph. Avouris,
{\it Conductance of Distorted Carbon Nanotubes},
cond-mat/9904083

\bibitem{Odintsov99}
A. A. Odintsov and Y. Tokura, 
{\it Contact phenomena in carbon nanotubes},
cond-mat/9906269;
K Esfarjani, A. A. Farajian, Y. Hashi and Y. Kawazoe 
%Electronic and transport properties of N-P doped nanotubes,
Appl. Phys. Lett. {\bf 74}, 79 (1999);
F. Leonard and J. Tersoff,
{\it Novel length scales in nanotube devices},
Preprint

\bibitem{footnote2}
One can also take the Fermi energy to lie off band center but this does
not play an important role in conveying the main points of this paper.

\bibitem{footnote1}
Note that if the nanotube-metal coupling is strong, then the 
transmission probability would reach its maximum by contacting just 
only a few layers along the length. Further increase in contact length
will not result in a monotonic increases in transmission with contact
length.

\end{thebibliography}
\end{document}